\documentclass[runningheads]{cl2emult}

\usepackage{makeidx}  
\usepackage{graphicx} 
\usepackage{subeqnar} 
\usepackage{multicol} 
\usepackage{cropmark} 
\usepackage{math}     
\makeindex            

\begin{document}
\title*{Variety of Stock Returns in Normal and Extreme Market Days: 
The August 1998 Crisis.}
\toctitle{Variety of Stock Returns in Normal and Extreme Market Days: 
The 1998 Crisis.}
%
%
\titlerunning{Variety of Stock Returns}
%
\author{Fabrizio Lillo\inst{1}
\and Giovanni Bonanno\inst{1}
\and Rosario N. Mantegna\inst{1,2}}
\authorrunning{Fabrizio Lillo et al.}
%
%
\institute{Istituto Nazionale per la Fisica della Materia,
Unit\`a di Palermo, Facolt\`a di Ingegneria, Viale delle
Scienze, I-90128 Palermo, Italia
\and Dipartimento di Fisica e Tecnologie Relative,
     Universit\`a di Palermo,
     Viale delle Scienze,
     I-90128 Palermo, Italia}

\maketitle              

\begin{abstract}
We investigate the recently introduced variety of a set of 
stock returns
traded in a financial market. This investigation is done by 
considering daily and intraday time horizons in a 15-day time period 
centered at the August 31st, 1998 crash of the S\&P500 index.
All the stocks traded at the NYSE during that period are considered in
the present analysis. We show that the statistical properties 
of the variety observed in analyses of daily returns also hold for 
intraday returns.
In particular the largest changes of the variety of the return
distribution turns out to be most localized at the opening or 
(to a less degree) at the closing of the market.   
\end{abstract}

\section{Introduction}
In recent years physicists started to model financial markets as
complex systems (Anderson et al. 1988) within their academic 
research activity. This triggered the interest
of a group of physicists towards the analysis and modeling of price 
dynamics in financial markets performed by using paradigms
and tools of statistical and theoretical physics 
(Li 1991,Mantegna 1991, Takayasu 1992, Bak 1993, Bouchaud 1994, 
Mantegna and Stanley 1995, Mantegna 1999b, Bouchaud et al. 2000b). 
One target of these researches is to implement a new stochastic 
model of price dynamics in 
financial markets which reproduces the statistical properties
observed in the time evolution of financial stocks
(Mantegna and Stanley 2000, Bouchaud and Potters 2000).
In the last few years physicists performed several empirical 
researches investigating the 
statistical properties of price and volatility time series 
of a single stock (or of an index) at different time
horizons (M\"uller et al. 1995, Mantegna and Stanley 1995, 
Lux 1996, Gopikrishnan et al. 1998). 
Such a kind of analysis does not take
into account any interaction of the considered financial stock 
with other stocks which are traded simultaneously in the same market.
It is known that the synchronous price returns time series of 
different stocks are pair correlated (Elton and Gruber 1995,
Campbell et al. 1997) and several
researches has been performed also by physicists in order to 
extract information from the correlation properties 
of a set of stocks (Mantegna 1999a, Laloux et al 1999, 
Plerou et al. 1999). A precise
characterization of collective movements in a financial market 
is of key importance in understanding the market dynamics and in 
controlling the risk associated to a portfolio of stocks. The 
present study presents some of the results obtained by our group 
about the collective behavior of an ensemble of 
stocks in normal and extreme days of market activity.  
This is done by discussing the main concepts recently
introduced and by presenting them by using a case study 
focused on the August 1998 crisis of the New York 
Stock Exchange (NYSE).

Some properties of the collective behavior of stocks traded 
simultaneously in a market are studied by considering the
ensemble properties of a set of stocks. Specifically, we 
investigate the stock returns of an 
ensemble of $n$ stocks simultaneously traded in a financial 
market at a given day. With this approach we
quantify what we have called the {\it variety} of the financial market 
at a given trading day (Lillo and Mantegna 2000a, Lillo and Mantegna 2000b). 
The variety provides statistical information about the amount of
different behavior observed in stock returns in a given ensemble 
of stocks at a given trading time horizon (in the present 
study, we obtain empirical results by investigating
time horizons ranging from one trading day down to 5 trading 
minutes). 
The shape and parameters characterizing the ensemble return distribution
are relatively stable during normal phases of the market activity 
whereas become time dependent in the periods subsequent to crashes.

The statistical properties of variety are sensitive to  the composition 
of the portfolio investigated (especially to the capitalization
of the considered stocks) and a simple model such as the
single-index model is not able to reproduce the statistical properties 
empirically observed. In this paper we present the results obtained by our group
about the synchronous analysis of the daily and high-frequency 
returns of all the stocks traded in the NYSE during a period of
time centered around a significant market crash. 
The time period selected is a 15 trading days period centered
at the August 31st, 1998 crisis. At this day the S\&P500
experienced a -6.79\% drop, the fourth biggest one-day crash
of the last 50 years. 

The paper is organized as follow. In Section 2 we illustrate  
the statistical properties of the daily variety. 
Section 3 is devoted to study in detail the intraday behavior
of the variety. In Section 4, we present a 
brief discussion of the obtained results.

\section{The Variety of an Ensemble of Stocks Simultaneously Traded}

For presentation purposes we first start our analysis with a one day time horizon and
we then consider a high-frequency analysis of a period of crisis.
The first investigation is performed by extracting the $n$ 
returns $R_i$ of the $n=2798$ stocks traded in the NYSE for each trading day $t$
of our database covering the period from August 20th to September 10th
1998.
The distribution of these returns $P_t(R)$ provides  information 
about the kind of activity occurring in the market at the 
selected trading day $t$ belonging to a period of high volatility 
such as the one of August and September 1998. A study covering an
11-year time period is published in ref. (Lillo and
Mantegna 2000b) where it has been shown that 
a customary profile of the ensemble return distribution 
exists for typical market days. However, this profile is
not observed during days of large absolute market averages
(Lillo and Mantegna 2000c). 

\begin{figure}[t]
\centering
\includegraphics[width=1.0\textwidth]{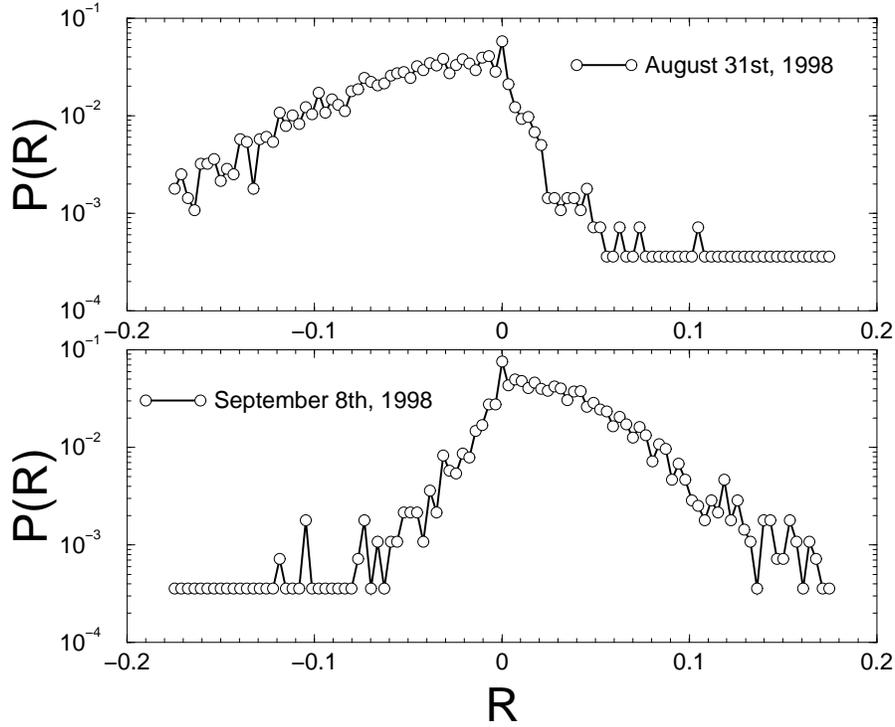}
\vspace{0.4cm}
\caption[]{Daily ensemble return distribution of all the equities 
traded in the New York Stock Exchange for the extreme trading 
days August 31st (top panel), and  September 8th, 1998 (bottom panel). 
The August 31st is the worst performing day of 1998
(-6.79\% of S\&P500), and the 8th September 1998 is the best 
rally day (+5.09\% of S\&P500). 
The skewness of the distribution is negative in crash (top), 
and positive in rally (bottom) days.}
\label{eps1}
\end{figure}

The time period investigated
in the present study is a period of large absolute market 
averages. Hence we expect that the ensemble return 
distribution bears the properties of asymmetry
observed for the first time in
ref. (Lillo and Mantegna 2000c).
Figure 1 shows the empirical return probability density 
function (pdf) 
for two days representative of extreme 
market days.
In this figure we show the interval of daily returns from 
$-20\%$ to $20\%$. The two distributions refer to the largest
drop and increase of the S\&P500 observed in the investigated
time period. Consistently with the results published 
in ref. (Lillo and Mantegna 2000c), we observe that the 
symmetry of the distribution is not conserved during extreme 
market days. Moreover, the stylized fact of observing a
negatively skewed distribution (top panel of Fig. 1) 
during a crash and a positively skewed distribution (bottom panel
of Fig. 1) during a rally is fully confirmed.

\begin{figure}[t]
\centering
\includegraphics[width=0.9\textwidth]{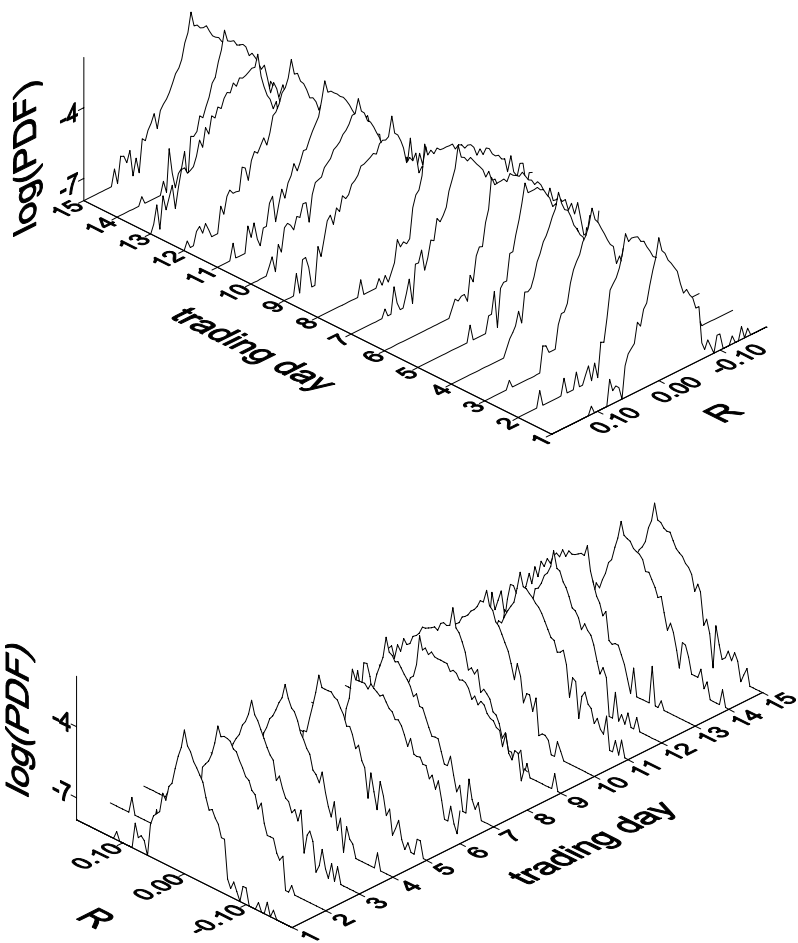}
\vspace{0.2cm}
\caption[]{Natural logarithm of the daily pdf of 2798 stock returns
traded in the NYSE. The investigated time period covers the 15 trading
days period starting at August 20th 1998. The crash of the 
investigated crisis occurs at August 31st (8th day of the figure),
while the most effective rally occurs at September 8th  (13th day 
of the figure). The 3D figures is provided from two different 
perspectives to direct observe both crashes (bottom panel) and rallies
(top panel). The return pdfs for trading days occurring before the
crash are characterized by an approximately symmetric 
profile. During crashes and rallies the return pdf becomes
asymmetric showing negative (8th day of the figure) and 
positive (13th day of the figure) skewness respectively.}
\label{fig2}
\end{figure}

In order to characterize the ensemble return
distribution at each day $t$ of the investigated period we determine
both the pdfs and the first 
two central moments for each of the $15$ investigated trading days. 
In Fig. 2 we show the 15 daily pdfs in a 3D plot both from the 
perspective of the crashes (bottom panel of the figure) and
from the perspective of the rallies (top panel of the figure).
During the trading days immediately before the crises  
(August 31st 1998 labeled as day 8 in the figure) we observe
a pdf of returns which is approximately symmetric and 
characterized by a non Gaussian shape. This shape is close
to the ``typical''
profile observed during normal market days (Lillo and Mantegna
2000b).

\begin{figure}[t]
\centering
\includegraphics[width=1.0\textwidth]{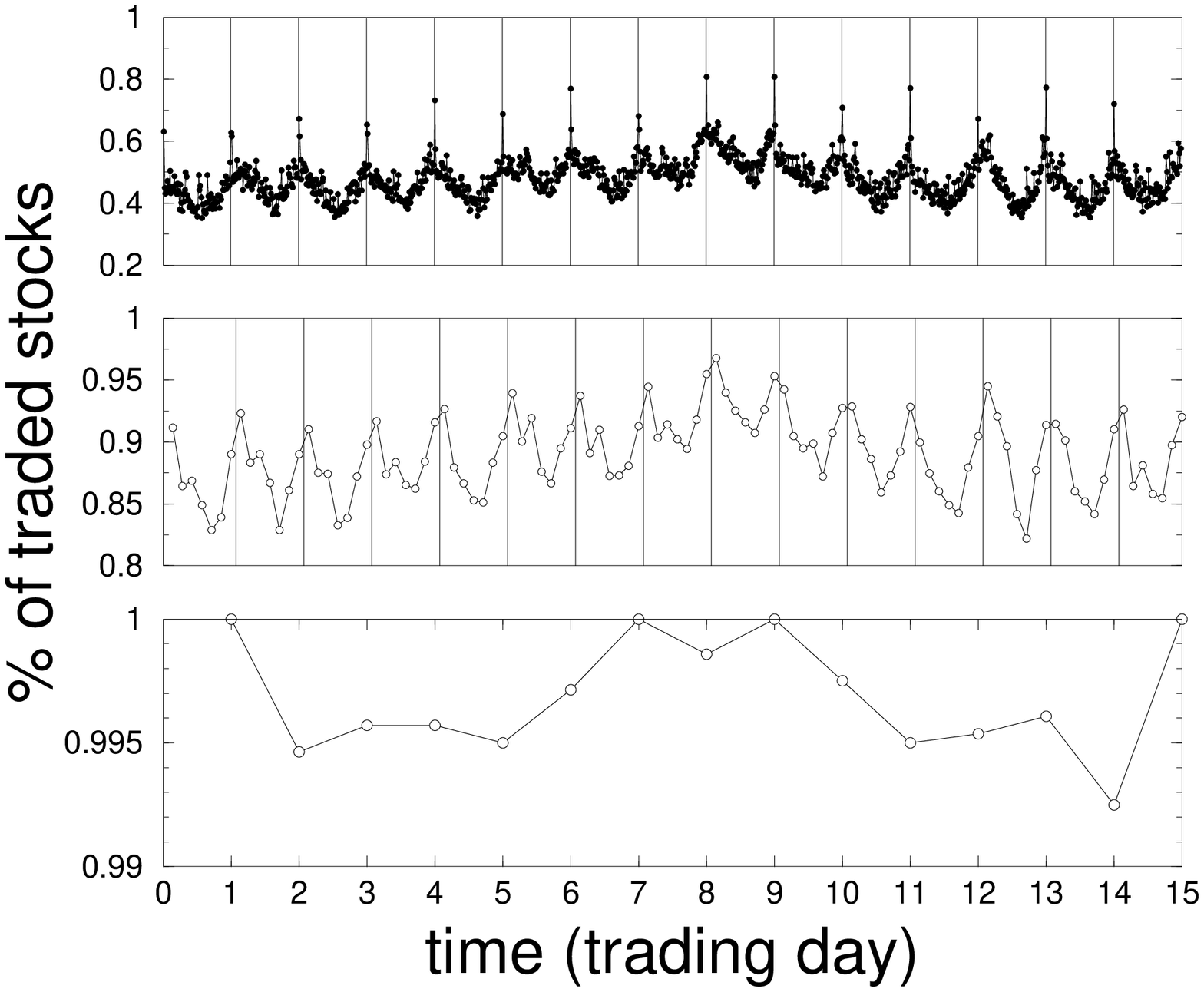}
\vspace{0.4cm}
\caption[]{Percentage of the 2798 stocks listed in the NYSE
having performed at least one transaction in a time window
of 5 minutes (top panel), 55 minutes (middle) panel and
390 minutes equal to one trading day (bottom panel). 
Vertical lines 
indicates the closing and opening of trading days.}
\label{fig3}
\end{figure}

The abrupt change of the shape and parameters of the pdf
occurring at the crisis and immediately after is 
documented in this figure.
The most extreme deformation of the symmetry and parameters of the 
pdf is observed at the 8th day and at the 13th day of of the 
investigated period. These days correspond to August 31st and September
8th 1998 which are two days when the S\&P500 experienced a
variation of -6.79\% and +5.09\% respectively.

\section{Intraday Behavior of the Variety During a Crisis}

The daily investigation summarized in Fig. 2
does not provide any information about 
the intraday behavior of the variety and of the ensemble
return distribution. We use the {\it Trade and Quote} 
database of NYSE to investigate the high frequency behavior
of the ensemble return distribution and of its 
mean and standard deviation (the variety). 
This is done by investigating the return of 2798 stocks
traded in the NYSE by using time horizons of 55 and
5 minutes.  

A problem experienced by investigating the returns
of an ensemble of stocks at high frequency is 
that not all the investigated stocks are traded
at each investigated time interval. The number of
stocks that are not traded becomes relevant at very short
time horizons. Before we perform our analysis we 
calculate the percentage of stocks traded for each investigated
time interval. The results are summarized in Fig. 3 where 
we show the percentage of stocks traded in the 
investigated time period by investigating 5 minutes
(top panel), 55 minutes (middle panel) and one 
trading day (bottom panel) time intervals. From the 
figure we note that at a 5 minutes time interval a
percentage between 40\%  and 60\% of traded stocks is
observed. This level of trading increases immediately after the 
opening when percentage of the order of 80\% of traded stocks are 
observed during the first 5 minutes of trading.
The level of trading increases to values between 80\% 
and 95\% when time intervals of 55 minutes are considered
(middle panel of the figure). This percentage further
increases to values close to 100\% when a one day time 
interval is considered (bottom panel of the figure).

To make our investigation at different time horizons 
consistent we performed all our analyses over the 
set of stocks that in each time horizon are {\it effectively}
traded. In other words, we include each stock in 
our analysis if during the time interval considered 
is traded at least one time. In our analysis we also 
consistently remove all the overnight returns. 

\begin{figure}[t]
\centering
\includegraphics[width=1.0\textwidth]{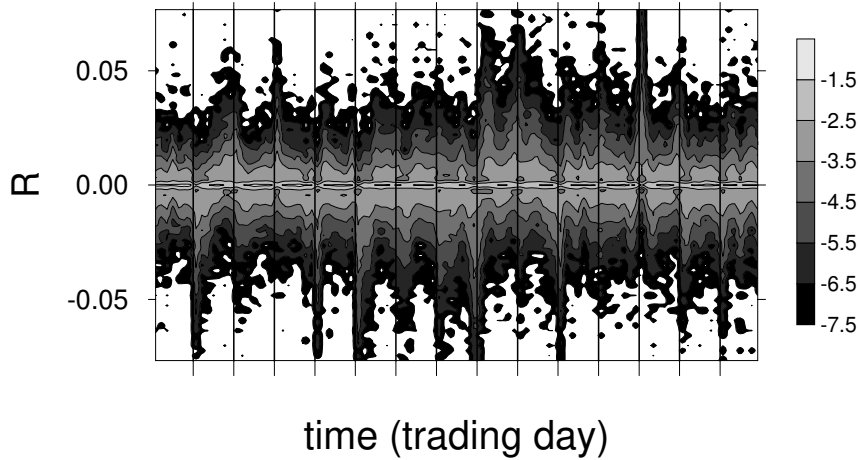}
\caption[]{Contour plot of the logarithm of the return pdf
for the 15-day investigated time period. The intraday 
time horizon used to compute the return pdf is 55 trading minutes.
The contour plot is obtained for equidistant intervals of 
the logarithmic probability density. The central brightest area of 
the contour plot corresponds to the most probable value. The 
darker regions correspond to less probable. The contour lines
are obtained by considering the natural logarithm of the return pdf. 
The numbers of the gray scale are given in this unit. Each vertical line
indicates the closing of a market day. After the 8th day 
(the crash day of August 31st) the contour plot becomes more distorted. 
The strongest distortion is observed immediately after the opening
and occasionally near the closing.}
\label{fig4}
\end{figure}

In Fig. 4 we show the contour plot of the 55 minutes 
time horizon ensemble return distribution. The figure
is drawn by using a gray scale. Each gray level refers
to an interval between two contour plots of the logarithm
of the ensemble pdf. Vertical lines are indicating the 
closing (and opening) of each trading day. The contour plot 
shows that the larger broadening, distortion and swing 
of the return pdf is observed close to the opening and 
closing of each trading day.

The end of summer 1998 was a period of high variety 
(see, for example, Fig. 4 of ref. (Lillo and Mantegna 2000b)).
Even in the presence of a generalized high level of
variety, Fig. 4 shows that after the end of August 
crisis, consistently with similar results observed
in different market periods (Lillo and Mantegna 2001), 
there is a relative increase of the variety during
the days immediately after the August 31st drop. 

To provide a more explicit tracking of the variety 
of the market observed during the selected time period,
we directly calculate it in parallel with the market 
average $\mu(t)$. Specifically, we consider the
average  and the standard  deviation defined as
\begin{eqnarray}
&&\mu (t)=\frac{1}{n}\sum_{i=1}^{n} R_i(t), \\
&&\sigma (t)= \sqrt{\frac{1}{n}\left(\sum_{i=1}^{n}
(R_i(t)-\mu(t))^2\right)},
\end{eqnarray}
where $n$ indicates the number of stocks effectively 
traded in the investigated period.

The mean of price returns $\mu(t)$ quantifies  the general trend
of  the market at day $t$.  The standard deviation $\sigma(t)$
is the variety of the market and gives a measure of the width 
of  the ensemble return distribution.
A large value of the variety $\sigma(t)$
indicates that different stocks are characterized by rather
different returns at day $t$.  In fact, in days of high variety,
some stocks perform great gains whereas others have great
losses. The mean and the standard
deviation of price returns are not constant and fluctuate in time.

\begin{figure}[t]
\centering
\includegraphics[width=1.0\textwidth]{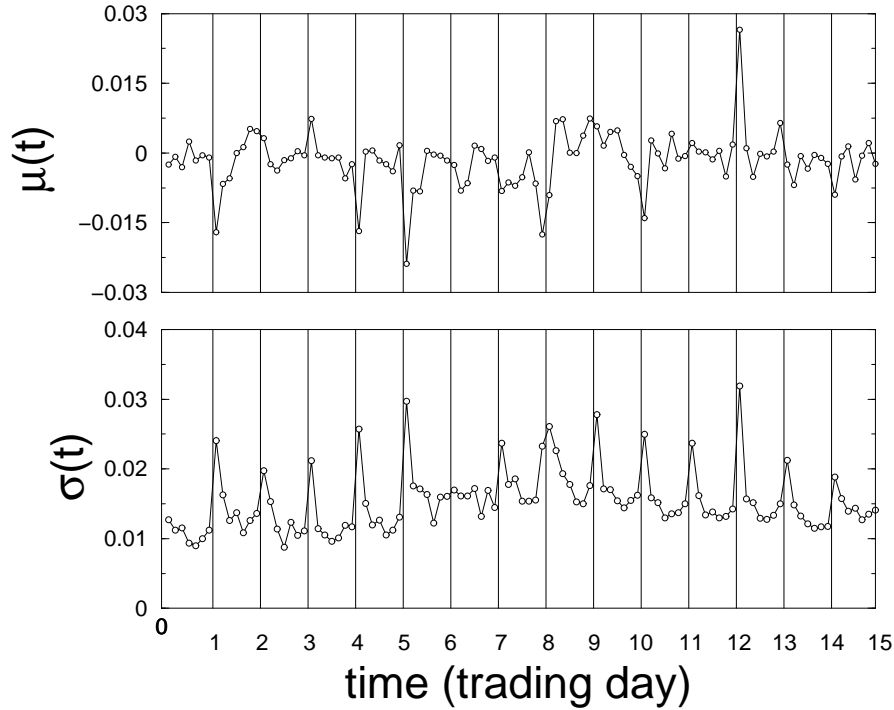}
\vspace{0.4cm}
\caption[]{Time series of the market average $\mu(t)$
(top panel) computed by using a 55 minutes time horizon.
The largest absolute values of the market averages are
occurring near the opening or closing of the market.
Each vertical line refers to the closing of a market day.
The first trading day is August 20th 1998 and the
8th trading day is the crash day of August 31st. 
In the bottom panel we show the time series of the variety 
$\sigma(t)$ determined under the same conditions as in the
top panel. Again the largest variety is observed near the
opening or closing of a market day. After the onset of the
crisis (August 31st) the average level of the variety increases 
in in the market.}
\label{fig5}
\end{figure}

In Fig. 5 we show the market average $\mu(t)$ (top panel) 
and the variety $\sigma(t)$ (bottom panel) computed
with a 55 minutes time horizon in the investigated period.
By using the same presentation scheme of Fig. 3, vertical lines 
indicates the closing and opening of trading days.
By inspecting Fig. 5 one proves quantitatively the relative
increases of the variety observed at the August 31st crisis 
and immediately after. There is also additional information
concerning the intraday localization of the moments of
highest variety. Spikes of variety are localized at the 
opening of the market and to a less degree near the closing 
of the market day. Large values of variety are associated
with large values of absolute market average. The relation
between these two variables has been worked out within the
framework of the single-index model (Lillo and Mantegna 2001). 

The single-index model (Elton and Gruber 1995, Campbell et al. 1997)
assumes that $R_i(t)$ can be written as:
\begin{equation}
R_i(t) = \alpha_i + \beta_i R_m(t) + \epsilon_i(t),
\end{equation}
where $\alpha_i$ is the expected value of the component 
of security $i$'s return that is independent of the market's
performance, $\beta_i$ is a coefficient usually close to unity,
$R_m(t)$ is the market factor and
$\epsilon_i(t)$ is called the idiosyncratic return, by construction
uncorrelated to the market.

\begin{figure}[t]
\centering
\includegraphics[width=1.0\textwidth]{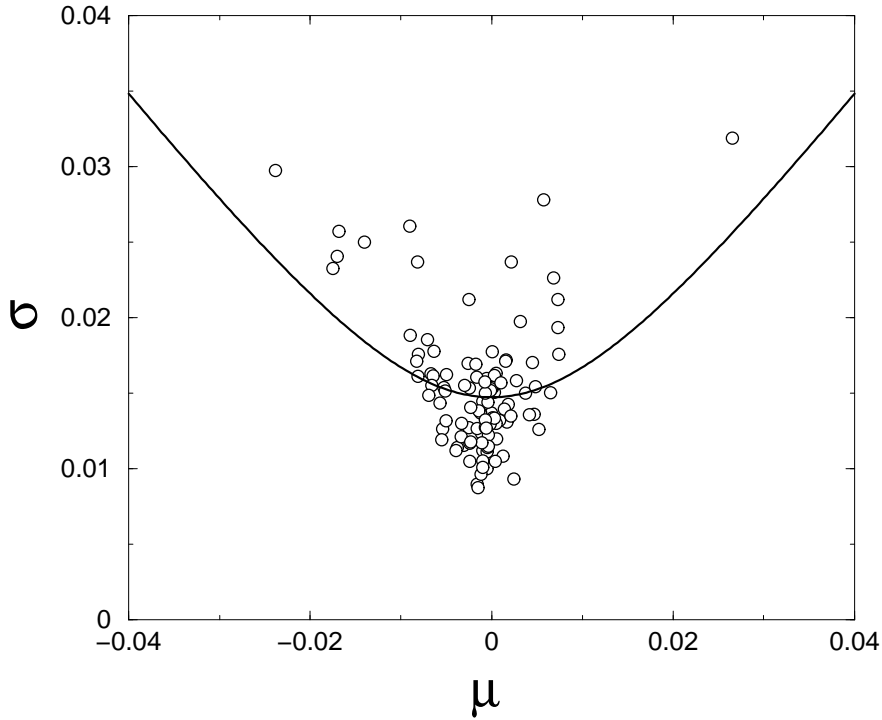}
\caption[]{Variety $\sigma(t)$ of the return pdf 
as a function of the market average $\mu(t)$ for each 55-minute
intraday time intervals of the 15-day investigated time period. 
Each circle refers to a 55-minute intraday time interval. 
The solid line is the theoretical prediction of Eq. (4) with 
the parameters detected by least square procedures with
the market average used as market factor. Specifically,
$<\epsilon_i^2(t)>=2.17 \cdot 10^{-4}$ and 
$(<\beta_i^2>-<\beta_i>^2)/<\beta_i>^2=0.6225$. The variety 
determined in the presence of large values of the 
absolute market average is underestimated by the 
single-index model.}
\label{fig6}
\end{figure}

Indeed, under the assumptions that $\alpha_i$ parameters 
can be neglected, the relation between the variety 
and the market average return is well approximated as
\begin{equation}
\sigma(t) \simeq \sqrt{<\epsilon_i^2(t)> 
+\frac{<\beta_i^2>-<\beta_i>^2}{<\beta_i>^2} \mu^2(t)}
\end{equation}
where $<\epsilon_i^2(t)>$ is the mean square
value of idiosyncratic terms and the symbol $< \cdot >$ 
indicates the average over all stocks $i$ of the considered parameter.

\begin{figure}[t]
\centering
\includegraphics[width=1.0\textwidth]{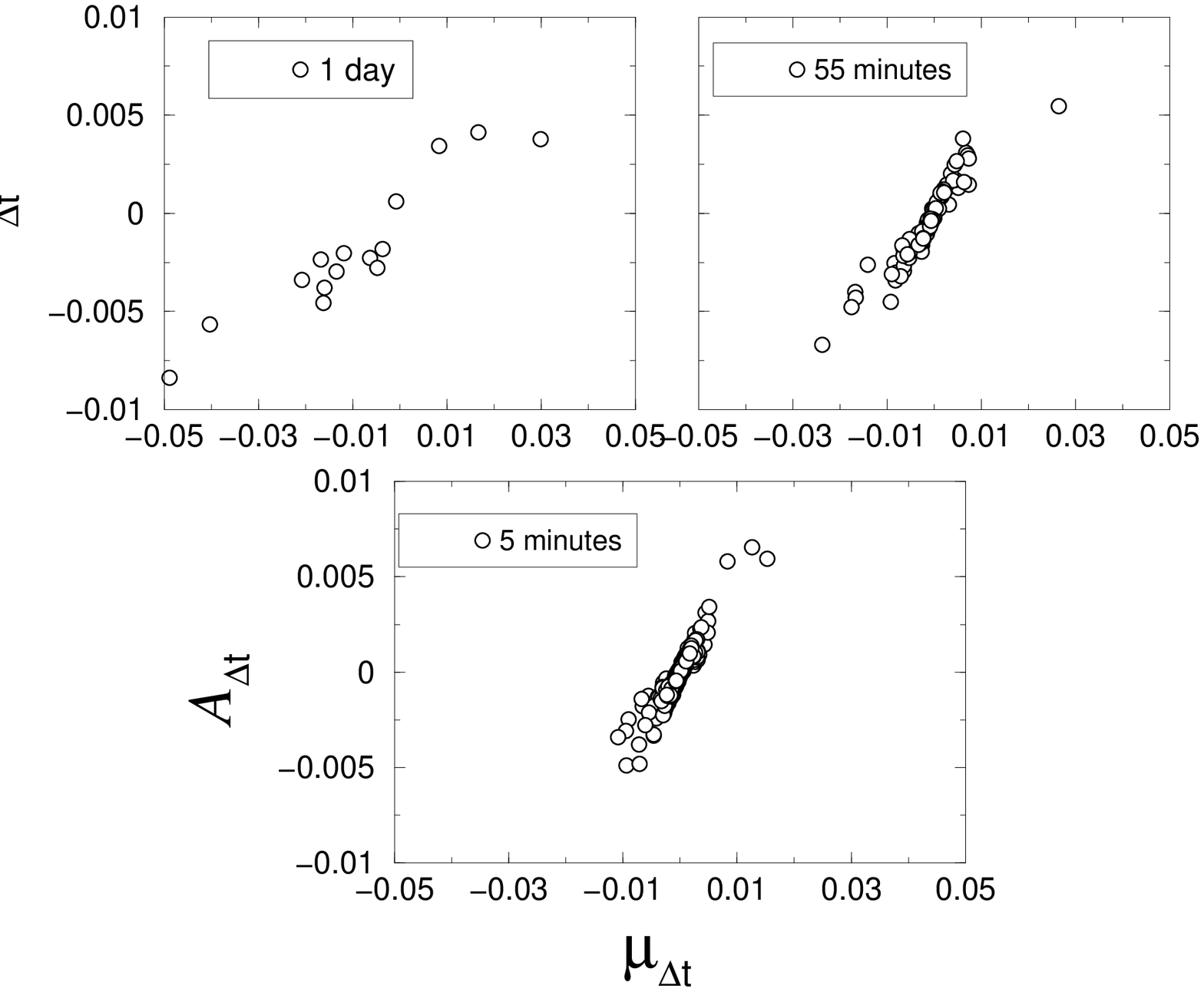}
\vspace{0.4cm}
\caption[]{Asymmetry $A_{\Delta t}$ measured as the mean minus the median 
of the return pdf as a function of the market average $\mu_{\Delta t}$ 
for different trading intervals of the investigated time period.
In the top left panel we show results obtained by
using a daily time horizon during the time period of the 
present investigation (from August 20th to September 10th 1998). 
A 55 minutes time horizon is used to obtain the top right 
panel, and the bottom panel is obtained by investigating 
the time period with a 5 minutes time horizon.
The results obtained for all three the time horizons
show that empirical results always cluster on a typical pattern.
The asymmetry in the return pdf is empirically detected down
to a time interval as short as 5 trading minutes.
This pattern assumes a sigmoid shape for longer time horizons.
It is worth noting that this empirical behavior cannot be 
modeled with a simple single-index model.}
\label{fig7}
\end{figure}

The single-index model explains the general relation between 
variety and market average. However the quantitative
comparison of the theoretical predictions of Eq. (4) 
with the empirical results is not satisfactory for the time 
intervals characterized by a large absolute
market average and variety.
This empirical observation is summarized in Fig. 6 where we present
for each 55-minute interval of our investigated time period 
(corresponding to each open circle) the variety versus the
market average. In the same figure, we also show the theoretical 
prediction of the single-index model of Eq. (4) as a solid line.
This theoretical prediction is obtained by using the market average
as a market factor.
From Fig. 6 it is evident that the single-index market 
underestimates the market variety empirically observed 
in the presence of large absolute values of the market average.

This is not the only empirical property which is not well 
described by the single-index model. In fact the 
single-index model also fails in describing the asymmetry
of the return pdf detected in the presence of large values of the
absolute market average (Lillo and Mantegna 2000c, Cizeau et al. 2001). 
In the following we investigate the daily and high frequency
asymmetry $A$ of the return pdf defined as
\begin{equation}
A_{\Delta t}(t)=\mu_{\Delta t}(t)-\mu^*_{\Delta t}(t)
\end{equation} 
where $\mu_{\Delta t}(t)$ is the market average at the time $t$ 
computed by using a time horizon $\Delta t$ and $\mu^*_{\Delta t}(t)$
is the median of the pdf at the same time and time horizon.
In our study $\Delta t$ is set equal to 1 trading day, 55 and 5 
trading minutes. 
We use an asymmetry measure based on the lowest possible
moments because the use of asymmetry parameters based on
higher moments (such as, for example, the skewness) would
provide an estimation heavily dependent on the most rare
events. With our choice of an asymmetry measure based on the 
mean and the median of the return pdf we obtain empirical
measure of the asymmetry which is statistically robust.
 
In Fig. 7 we show the asymmetry $A_{\Delta t}$ for daily,
55 and 5 minutes time horizons. For daily returns (top
left panel), we observe that the sigmoid
shape of the asymmetry curve, already detected in 
(Lillo and Mantegna 2000c), 
is also observed in the investigated 15-day time interval 
of the 1998 crisis. By investigating shorter time horizons,
it is worth noting that the value of the 
asymmetry $A$ also depends on the absolute value of the market 
average $\mu$ when intraday time horizons are used.
In fact the top right panel (55 minutes time horizon)
and the bottom panel (5 minutes time horizon) of Fig. 7 show
that the change in the asymmetry of the return pdf occurs
down a time horizon as short as 5 trading minutes.

\section{Discussion}
The variety of an ensemble of stocks simultaneously traded 
in a financial market provides a direct tool allowing
to monitor the overall behavior of a market in a simple and 
direct way. 

In spite of its simplicity (variety can be straightforwardly
estimated from market data), the variety is providing 
information about the status of the market which cannot 
be taken into account by widely used market models such as 
the single-index model. The variety (and indeed the 
return pdf) can be tracked very frequently down to 
very short time horizons (in the present study we reached
the 5 minutes time horizon) providing information 
on the ability of a single-index model to describe the 
dynamics of an ensemble of stocks simultaneously traded. 

{\bf Acknowledgements} -- The authors thank INFM and MURST 
for financial support. This work is part of the FRA-INFM 
project {\it Volatility in financial markets}. G. Bonanno
and F. Lillo acknowledge FSE-INFM for their fellowships.

\clearpage
\addcontentsline{toc}{section}{Index}
\flushbottom
\printindex


\begin{thebibliography}{7}
%
\addcontentsline{toc}{section}{References}


\bibitem{Anderson1988}
Anderson, P.~W., Arrow, K.~J., Pines, D., (eds) (1988) 
The Economy as an Evolving Complex System. Addison-Wesley, 
Redwood City

\bibitem{Bak1993}
Bak, P., Chen, K., Scheinkman, J., Woodford, M. (1993)
Aggregate Fluctuations from Independent Sectoral Shocks: Self-Organized
Criticality in a Model of Production and Inventory Dynamics. 
Ricerche Economiche {\bf 47}, 3--30

\bibitem{Bouchaud1994}
Bouchaud, J.-P., Sornette, D. (1994)
The Black \& Scholes Option Pricing Problem in 
Mathematical Finance: Generalization and Extensions for a
Large Class of Stochastic Processes.
J. Phys. I France {\bf 4}, 863--881

\bibitem{Bouchaud2000a}
Bouchaud J.-P., Potters, M. (2000) Theory of financial 
risk. Cambridge University Press, Cambridge, UK

\bibitem{Bouchaud2000b}
Bouchaud, J.-P., Lauritsen, K., Alstrom, P., (eds) (2000b)
Proceedings of the International Conference on  Application 
of Physics in Financial Analysis. Int. J. Theor. Appl. 
Finance {\bf 3}, 309--608

\bibitem{Campbell1997}
Campbell, J.~Y., Lo, A.~W., MacKinlay, A.~C. (1997)
The Econometrics of Financial Markets. Princeton 
University Press, Princeton.

\bibitem{Cizeau2001}
Cizeau, P.,  M. Potters, M., Bouchaud, J.-P. (2001)
Correlation structure of extreme stock returns.
Quantitative Finance {\bf 1}, 217-222


\bibitem{Elton1995}
Elton, E.~J., Gruber, M.~J. (1995) Modern Portfolio Theory 
and Investment Analysis. J. Wiley \& Sons, New York

\bibitem{Gopikrishnan1998}
Gopikrishnan, P., Meyer, M., Amaral, L.~A.~N., Stanley, H.~E. (1998)
Inverse cubic law for the distribution of stock price
variations.
Eur. Phys. J. B {\bf 3} 139-140. 

\bibitem{Laloux1999}
Laloux, L., Cizeau, P., Bouchaud, J.-P., Potters, M. (1999)
Noise Dressing of Financial Correlation Matrices. 
Phys. Rev. Lett. {\bf 83}, 1467--1470

\bibitem{Li1991}
Li, W., (1991)
Absence of $1/f$ Spectra in Dow Jones Daily Average. 
Int'l J. Bifurcations and Chaos {\bf 1}, 583--597

\bibitem{Lillo2000a}
Lillo, F., Mantegna, R.~N. (2000a)
Statistical Properties of Statistical Ensembles of Stock Returns.
Int. J. Theor. Appl. 
Finance {\bf 3}, 405--408

\bibitem{Lillo2000b}
Lillo, F., Mantegna, R.~N. (2000b)
Variety and Volatility in Financial Markets.
Phys. Rev. E {\bf 62} 6126-6134

\bibitem{Lillo2000c}
Lillo, F., Mantegna, R.~N. (2000c)
Symmetry alteration of ensemble return distribution in 
crash and rally days of financial market. 
Eur. Phys. J. B {\bf 15} 603-606. 

\bibitem{Lillo2001}
Lillo, F., Mantegna, R.~N. (2001)
Empirical properties of the variety of a financial 
portfolio and the single-index model,
Eur. Phys. J. B, in press 

\bibitem{Lux1996}
Lux, T. (1996)
The stable Paretian hypothesis and the frequency of large 
returns: an examination of major German stocks,
Applied Financial Economics {\bf 6} 463-475. 

\bibitem{Mantegna1991}
Mantegna, R.~N. (1991) 
L\'evy Walks and Enhanced Diffusion in Milan Stock Exchange.
Physica A {\bf 179}, 232--242


\bibitem{Mantegna1995}
Mantegna, R.~N., Stanley, H.~E. (1995) 
Scaling Behaviour in the Dynamics of an Economic Index.
Nature {\bf 376}, 46--49

\bibitem{Mantegna1999a}
Mantegna, R.~N. (1999a) 
Hierarchical Structure in Financial Markets.
Eur. Phys. J. B {\bf 11}, 193--197


\bibitem{Mantegna1999b}
Mantegna, R.~N., (ed) (1999b) Proceedings of the International Workshop on 
Econophysics and Statistical Finance. Physica A {\bf 269}, 1--188

\bibitem{Mantegna2000}
Mantegna, R.~N., Stanley, H.~E. (2000) An Introduction 
to Econophysics: Correlations and Complexity in Finance. 
Cambridge University Press, Cambridge, UK

\bibitem{Muller1995}
M\"uller, U.~A., Dacorogna, M.~M., Olsen, R.~B., Pictet, O.~V.,
Schwarz, M. (1995)
Statistical Study of Foreign Exchange Rates, Empirical Evidence
of a Price Change Scaling Law and Intraday Analysis.
J. Banking and Finance {\bf 14} 1189-1208. 

\bibitem{Plerou1999}
Plerou, V., Gopikrishnan, P., Rosenow, B., Amaral L.~A.~N., 
Stanley, H.~E. (1999)
Universal and Nonuniversal Properties of Cross 
Correlations in Financial Time Series. 
Phys. Rev. Lett. {\bf 83}, 1471-1474

\bibitem{Takayasu1992}
Takayasu, H., Miura, H., Hirabayashi, T., Hamada, K. (1992) 
Statistical Properties of Deterministic Threshold 
Elements -- The Case of Market Price.
Physica A {\bf 184}, 127--134


\end{thebibliography}
\end{document}